# A Proposal on Quantum Histone Modification in Gene Expression

Liaofu Luo*

Laboratory of Theoretical Biophysics,  Faculty of Physical Science and Technology,

Inner Mongolia University,  Hohhot 010021,  China

*Email address: lolfcm@mail.imu.edu.cn

**Abstract**
A quantum mechanical model on histone modification is proposed. Along with the methyl / acetate or other groups bound to the modified residues the torsion angles of the nearby histone chain are supposed to participate in the quantum transition cooperatively. The transition rate $W$ is calculated based on the non-radiative quantum transition theory in adiabatic approximation. By using $W$'s the reaction equations can be written for histone modification and the histone modification level can be calculable from the equations, which is decided by not only the atomic group bound to the modified residue, but also the nearby histone chain.  The theory can explain the mechanism for the correlation between a pair of chromatin markers observed in histone modification. The temperature dependence and the coherence-length dependence of histone modification are deduced. Several points for checking the proposed theory and the quantum nature of histone modification are suggested as follows: 1, The relationship between ln$W$ and $1/T$ is same as usual protein folding. The non-Arhenius temperature dependence of the histone modification level is predicted. 2, The variation of histone modification level through point mutation of some residues on the chain is predicted since the mutation may change the coherence-length of the system. 3, Multi-site modification obeys the quantum superposition law and the comparison between multi-site transition and single modification transition gives an additional clue to the testing of the quantum nature of histone modification.

## 1 Introduction
Histone modification levels are predictive for gene expression [1][2]. Histone modification profiles are also predictive for tissue/cell-type specific expression of both protein-coding and microRNA genes [3]. It was indicated that only multiple different covalent modifications at the terminus of histones can play their role cooperatively or antagonistically in gene expression (the so-called "histone code" hypothesis [4]). The combinatorial patterns of histone modifications were studied in literatures [5][6] and their causal and combinatorial relationships were inferred by Bayesian network method [7]. It is known that the chromatin markers are essentially bivalent-like. A pair of histone modifications generally involves an active and a repressive marker. For example, it was reported that quantitative models involving H3K4me3 (active) and H3K79me1 (repressive) are the most predictive of the expression levels in low CpG content promoters, whereas high CpG content promoters require H3K27ac (active) and H4K20me1 (repressive) [2]. How to understand the mechanism for the correlation between a pair of covalent modifications?

In a recent theory the protein folding was regarded as a quantum transition between torsion states on polypeptide chain [8][9]. The importance of torsion state can be looked as follows: as a multi-atom system, the conformation of a protein is fully determined by bond lengths, bond angles, and torsion angles (dihedral angles), among which the torsion angles are most easily changed even at room temperature and usually assumed as the main variables of protein conformation. Simultaneously, the torsion potential generally has several minima the transition between which is responsible for the conformational change. All torsion modes between contact residues are assumed to participate in the quantum transition coherently in protein folding. The successes of the theory imply that many cooperative phenomena observed in molecular biology may be of quantum origin and the quantum coherence of torsional transitions may be responsible for the mechanism of cooperative motion of conformational-electronic system. In the following we will propose a mathematical theory to explain the long-distance correlation between bivalent-like chromatin markers.

The Histone Modification (HM) can be modeled as follows: Lysine, arginine and serine (threonine) on histone tail can be acetylated, methylated or phosphorylated [10]. Suppose a pair of modifications occurring at sites $X_a$ and $X_r$. Along with the atomic groups (one two or three methyl groups or one acetate group, etc) bound to the modified residues the torsion angles $\{\theta\}$ of histone chain between $X_a$ and $X_r$ (possibly including torsion angles on the nearby histone modifying enzymes) participate in the quantum transition cooperatively. The dynamical variables of the system are denoted by $(\theta, x)$ where $x=\{x_a, x_r\}$ represents the coordinate of atomic groups bound at site $X_a$ and $X_r$ respectively. We shall study the histone modification problem based on quantum conformational dynamics and discuss the experimental implications of the theoretical results.

## 2 Calculation of the histone modification rate

The wave function of the HM system $M(\theta, x)$ satisfies

$$(H_1(\theta, \frac{\partial}{\partial \theta}) + H_2(\theta, x, \nabla))M(\theta, x) = EM(\theta, x) \tag{1}$$

$H_1(\theta, \frac{\partial}{\partial \theta})$ is the torsion Hamiltonian $H_1(\theta, \frac{\partial}{\partial \theta})$ including kinetic energy term $\sum(-\frac{\hbar^2}{2I_j}\frac{\partial^2}{\partial \theta_j^2})$ ($I_j$ is the inertial moment of the $j$-th mode) and torsion potential term and $H_2(\theta, x, \nabla)$ describes the motion of atomic groups interacting with the modified residues. The latter is dependent of $\{\theta\}$ as parameters. The equation (1) is difficult to solve. However, under the assumption that the velocity of atomic groups in binding process is much higher than the torsion vibration of histone chain the adiabatic approximation can be used. In this approximation the wave function of HM is expressed as

$$M(\theta, x) = \psi(\theta)\varphi(x, \theta) \tag{2}$$

and these two factors satisfy

$$H_2(\theta, x, \nabla)\varphi_\alpha(x, \theta) = \varepsilon^\alpha(\theta)\varphi_\alpha(x, \theta) \tag{3}$$

$$\{H_1(\theta, \frac{\partial}{\partial \theta}) + \varepsilon^\alpha(\theta)\}\psi_{kn\alpha}(\theta) = E_{kn\alpha}\psi_{kn\alpha}(\theta) \tag{4}$$

here $\alpha$ denotes the atomic-group state, and $(k, n)$ refer to the torsional conformation- and vibration-state of histone chain, respectively.

Because Eq (2) is not a rigorous eigenstate of Hamiltonian $H_1 + H_2$, there exist transitions between adiabatic states that result from the off–diagonal elements [11]

$$\int M^+_{k'n'\alpha'}(H_1 + H_2)M_{kn\alpha}d\theta d^3x = E_{kn\alpha}\delta_{kk'}\delta_{nn'}\delta_{\alpha\alpha'} + \langle k'n'\alpha' | H' | kn\alpha \rangle \tag{5}$$

$$\langle k'n'\alpha' | H' | kn\alpha \rangle = \int \psi^+_{k'n'\alpha'}(\theta)\{-\sum_j \frac{\hbar^2}{2I_j}\int \varphi^+_{\alpha'}(\frac{\partial^2 \varphi_\alpha}{\partial \theta_j^2} + 2\frac{\partial \varphi_\alpha}{\partial \theta_j}\frac{\partial}{\partial \theta_j})d^3x\}\psi_{kn\alpha}(\theta)d\theta \tag{6}$$

Here $H'$ is a Hamiltonian describing the HM transition. We see that the atomic-group transition is always accompanied with the torsion transition from $\psi_{kn\alpha}$ to $\psi_{k'n'\alpha'}$.

The $\theta$ dependence of atomic-group wave function $\varphi_\alpha(x, \theta)$ can be deduced by the perturbation method,

$$H_2(\theta, x, \nabla) = H_2(\theta_0, x, \nabla) + \sum_j (\frac{\partial H_2(\theta, x, \nabla)}{\partial \theta_j})_0(\theta_j - \theta_{j0})$$

$$\equiv H_2(\theta_0, x, \nabla) + \sum_j h^{(j)}(x, \nabla)(\theta_j - \theta_{j0})$$

$$h^{(j)}(x, \nabla) = (\frac{\partial H_2(\theta, x, \nabla)}{\partial \theta_j})_0 \tag{7}$$

$$\varphi_\alpha(x, \theta) = \varphi_\alpha(x, \theta_0) + \sum_j (\theta_j - \theta_{j0})\sum_{\beta \neq \alpha} \frac{h^{(j)}_{\beta\alpha}}{\varepsilon_0^\alpha - \varepsilon_0^\beta}\varphi_\beta(x, \theta_0) \tag{8}$$

Here

$$H_2(\theta, x, \nabla) = H_{2a}(\theta, x_a, \nabla_a) + H_{2r}(\theta, x_r, \nabla_r)$$

$$h^{(j)}(x, \nabla) = h_a^{(j)}(x_a, \nabla_a) + h_r^{(j)}(x_r, \nabla_r)$$

And $\alpha = (\alpha_a, \alpha_r), \beta = (\beta_a, \beta_r)$;

$$\varepsilon_0^\alpha = \varepsilon_0^{\alpha_a} + \varepsilon_0^{\alpha_r}, \quad \varepsilon_0^\beta = \varepsilon_0^{\beta_a} + \varepsilon_0^{\beta_r}$$

are eigenvalues of $H_2(\theta_0, x, \nabla)$ and corresponding eigenfunctions

$$\varphi_\alpha(x, \theta_0) = \varphi_{\alpha_a}(x_a, \theta_0)\varphi_{\alpha_r}(x_r, \theta_0)$$

$$\varphi_\beta(x, \theta_0) = \varphi_{\alpha_\beta}(x_\beta, \theta_0)\varphi_{\alpha_r}(x_r, \theta_0)$$

Inserting (8) into (6), only the second term is retained and one has

$$\langle k'n'\alpha' | H' | kn\alpha \rangle$$

$$= \sum_j \frac{i\hbar}{\sqrt{I_j}} a_{\alpha'\alpha}^{(j)} \int \psi^+_{k'n'\alpha'}(\theta)\frac{\partial}{\partial \theta_j}\psi_{kn\alpha}(\theta)d\theta \tag{9}$$

$$a^{(j)}_{\alpha'\alpha} = a^{(j)}_{\alpha'_a \alpha_a} + a^{(j)}_{\alpha'_r \alpha_r}$$

$$a^{(j)}_{\alpha'_i \alpha_i} = \frac{i\hbar}{I_j^{1/2}} \int \varphi^+_{\alpha'_i}(x_i, \theta_0) \sum_{\beta_i} \frac{h^{(j)}_{i,\beta_i \alpha_i}}{\varepsilon_0^{\alpha_i} - \varepsilon_0^{\beta_i}} \varphi_{\beta_i}(x_i, \theta_0) d^3 x_i \quad (i=a,r)$$

(10)

In the following we shall assume that the torsion potential can be approximated by a harmonic potential with parameter $\{\omega_j\}$ in the initial state and $\{\omega_j'\}$ in the final state. After thermal average over the initial states and summation over final states we have the HM transition rate

$$W = \frac{2\pi}{\hbar} \sum_{\{n\}} |\langle k'n'\alpha' | H' | kn\alpha \rangle|^2 B(\{n\},T) \rho_E \tag{11}$$

$$B(\{n\},T) = \prod_j e^{-n\beta\hbar\omega_j}(1 - e^{-\beta\hbar\omega_j}) \tag{12}$$

$$(\beta = \frac{1}{k_B T})$$

$B(\{n\},T)$ denotes the Boltzmann factor and $\rho_E$ means state density. Inserting (9)(10) into (11) it leads to

$$W = W_{dia} + W_{ndi} \tag{13}$$

$$W_{dia} = 2\pi\hbar \{ \sum_n \sum_j \frac{1}{I_j} |a^{(j)}_{\alpha'\alpha}|^2 \left| \int \psi^+_{k'n'\alpha'}(\theta) \frac{\partial}{\partial \theta_j} \psi_{kn\alpha}(\theta) d\theta \right|^2 B(\{n\},T) \rho_E \tag{14}$$

$$W_{ndi} = 2\pi\hbar \{ \sum_n \sum_{j \neq l} \frac{(a^{(j)}_{\alpha'\alpha})^* a^{(l)}_{\alpha'\alpha}}{\sqrt{I_j I_l}} \times$$

$$(\int \psi^+_{k'n'\alpha'}(\theta) \frac{\partial}{\partial \theta_j} \psi_{kn\alpha}(\theta) d\theta)^* \int \psi^+_{k'n'\alpha'}(\theta) \frac{\partial}{\partial \theta_l} \psi_{kn\alpha}(\theta) d\theta \} B(\{n\},T) \rho_E$$

(15)

The diagonal part is denoted as $W_{dia}$ (diagonal) and the non-diagonal as $W_{ndi}$. When $\omega_j = \omega_j'$, by using the phonon annihilation/production operator

$$\xi_j = (\frac{I_j \omega_j}{2\hbar})^{1/2} (\theta_j + \frac{1}{I_j \omega_j} \frac{\partial}{\partial \theta_j}), \quad \xi_j^+ = (\frac{I_j \omega_j}{2\hbar})^{1/2} (\theta_j - \frac{1}{I_j \omega_j} \frac{\partial}{\partial \theta_j})$$

(16)

we have

$$W_{dia} = \frac{\pi}{\hbar} \sum_j |a^{(j)}_{\alpha'\alpha}|^2 (\{(\overline{n}_j + 1) I_V(p-1) + \overline{n}_j I_V(p+1)\}$$

$$+ Q_j \{(6\overline{n}_j^2 + 6\overline{n}_j + 1) I_V(p) - 2(\overline{n}_j + 1)(2\overline{n}_j + 1) I_V(p-1)$$

$$- 2\overline{n}_j (2\overline{n}_j + 1) I_V(p+1) + (\overline{n}_j + 1)^2 I_V(p-2) + \overline{n}_j^2 I_V(p+2)\})$$

(17)

with

$$I_V(p) = \frac{1}{\sqrt{2\pi}} \exp(\frac{\Delta E}{2k_B T})(\sum Z_j)^{-\frac{1}{2}} \exp(-\frac{p^2}{2\sum Z_j})$$

$$I_V(p \pm 1) = \frac{1}{\sqrt{2\pi}} \exp(\frac{\Delta E}{2k_B T})(\sum Z_j)^{-\frac{1}{2}} \exp(-\frac{(p \pm 1)^2}{2\sum Z_j})$$

$$I_V(p \pm 2) = \frac{1}{\sqrt{2\pi}} \exp(\frac{\Delta E}{2k_B T})(\sum Z_j)^{-\frac{1}{2}} \exp(-\frac{(p \pm 2)^2}{2\sum Z_j}) \tag{18}$$

$$p = \sum \frac{\delta E_j}{\hbar \omega_j} = \Delta E / \hbar \overline{\omega} \quad \Delta E = \sum \delta E_j \tag{19}$$

$$Q_j = \frac{I_j \omega_j (\delta \theta_j)^2}{2\hbar} \tag{20}$$

$$Z_j = (\delta \theta_j^2) \frac{k_B T}{\hbar^2} I_j \tag{21}$$

$$\overline{n}_j = (e^{(\hbar \omega_j / k_B T)} - 1)^{-1} \simeq \frac{k_B T}{\hbar \omega_j} \tag{22}$$

Here $\delta \theta_j$ is the angular displacement and $\delta E_j$ the energy gap between the initial and final states for the $j$-th harmonic mode. In the above deduction the asymptotic formula for Bessel function [12]

$$e^{-z} J_p(z) = (2\pi z)^{-1/2} \exp(-p^2 / 2z) \quad \text{for} \quad z \gg 1 \tag{23}$$

and the recursion formula [8]

$$\sum_{p_1} \exp(-z_1) J_{p_1}(z_1) \exp(-z_2) J_{p-p_1}(z_2)$$

$$= \frac{1}{\sqrt{2\pi}} \frac{1}{\sqrt{z_1 + z_2}} \exp\{-\frac{p^2}{2(z_1 + z_2)}\} \tag{24}$$

have been used. The above results can be generalized to the case of non-equal frequencies $\omega_j \neq \omega_j'$ through the replacement of $\Delta E$ by

$$\Delta G = \Delta E + \sum_j \frac{1}{\beta} \ln \frac{\omega_j}{\omega_j'} \tag{25}$$

in (18)(19) [8].

The net variation $p$ of phonon number in torsion transition is much larger than 1 (see Eq 19,

$\Delta E \gg \hbar\bar{\omega}$) due to the number $N$ of torsion modes cooperatively participating in the transition not small.   Considering $I_V(p) \cong I_V(p\pm 1) \cong I_V(p\pm 2)$ Eq (17) can be simplified to

$$W_{dia} = \frac{\pi}{\hbar} \sum_j |a_{\alpha'\alpha}^{(j)}|^2 \{(2\bar{n}_j + 1) I_V(p)\}$$
$$\cong \frac{2\pi}{\hbar^2 \bar{\omega}'} N\bar{a}^2 k_B T I_V \qquad (26)$$

where $\bar{a}^2$ means the average of $|a_{\alpha'\alpha}^{(j)}|^2$ over $j$ and

$$I_V = \frac{1}{\sqrt{2\pi}} \exp(\frac{\Delta G}{2k_B T})(\sum Z_j)^{-\frac{1}{2}} \exp(-\frac{(\Delta G)^2}{2(\hbar\bar{\omega})^2 \sum Z_j}) \qquad (27)$$

$$\Delta G \cong \Delta E + Nk_B T \ln\frac{\bar{\omega}}{\bar{\omega}'} \qquad (28)$$

$$\sum Z_j \cong \frac{1}{\hbar^2} N I_0 (\delta\theta)^2 k_B T \qquad (29)$$

where $(\delta\theta)^2$ is the average of $\delta\theta_j^2$ over $j$, and $\bar{\omega}$, $\bar{\omega}'$ are the average frequency of initial and final state respectively and $I_0$ the average inertial moment.

The calculation of non-diagonal term $W_{ndi}$ is not easy. Following Sarai & Kakitani [13] one has

$$W_{ndi} = \sum_{j\neq l} \frac{\pi}{\hbar}(a_{\alpha'\alpha}^{(j)})^* a_{\alpha'\alpha}^{(l)} \sqrt{Q_j Q_l}$$
$$\{(2\bar{n}_j + 1)(2\bar{n}_l + 1)H(0) + 2(\bar{n}_j + 1)\bar{n}_l H(\hbar\omega_j - \hbar\omega_l)$$
$$-2(2\bar{n}_j + 1)(\bar{n}_l + 1)H(\hbar\omega_l) - 2\bar{n}_j(2\bar{n}_l + 1)H(-\hbar\omega_j)$$
$$+(\bar{n}_j + 1)(\bar{n}_l + 1)H(\hbar\omega_j + \hbar\omega_l) + \bar{n}_j\bar{n}_l H(-\hbar\omega_j - \hbar\omega_l)\} \qquad (30)$$

Here $H(x)$ is a function dependent of

$$p(\{n_j\}) = (\hbar\omega_s)^{-1}\{\Delta E - x - \sum_{j\neq s}\bar{n}_j \hbar\omega_j\} \qquad (31)$$

where $\omega_s$ is taken as the lowest frequency of all $N$ modes. As $\Delta E \gg \hbar\omega_j$, $p(\{n_j\})$ does not depend on $x$ and all $H(x)$ in Eq (30) are same.   Then we have

$$W_{ndi} \cong 0 \qquad (32)$$

as the difference of $a_{\alpha'\alpha}^{(j)}$ and $Q_j$ in different modes can be neglected. So the total rate $W$ is given by diagonal part, Eq (26).

## 3  Experimental implications of theoretical results

3.1   The gene expression is regulated by histone modification at transcriptional level. Suppose a pair of atomic groups g1 and g2 are bound to the histone tail H of a given promoter. The reaction equation can be written as

$$H_u + g1 + g2 + E \xrightarrow{W} H_m + E \qquad (33)$$

where $H_u$ and $H_m$ denote unmodified and modified histone respectively and E the histone modifying enzyme. A set of equations like (33) determines the concentration of $\{H_m\}$ （histone modification level）and therefore gives the gene expression value for each promoter. In the present quantum theory the histone modification level is decided by not only the atomic-group binding, but also the histone chain. The rate $\{W\}$ can be calculated through (13)(26) and (32). They can be compared with the experimental data on histone modification and the quantitative law of gene expression can be searched along the line.

3.2 The temperature dependence of histone modification is proposed in the present theory. From Eqs (26) to (29) the relation of the rate with temperature $T$ and torsion mode number $N$ can easily be deduced. Since $\Delta E$ is the sum of the energy gaps of $N$ torsion modes between initial and final states one assumes

$$\Delta E = bN^\gamma \quad (0 \le \gamma \le 1) \tag{34}$$

Inserting Eq (34) into (26) to (29) and denoting

$$c = 2k_B I_0 \bar{\omega}^2 (\delta\theta)^2 \tag{35}$$

we obtain

$$d\ln W = d\ln N + d\ln T + d\ln I_V$$
$$= F_N dN + F_T dT \tag{36}$$

$$\frac{\partial \ln W(T,N)}{\partial N} = F_N = \frac{1}{2N} + \frac{b\gamma}{2k_B T} N^{\gamma-1} - \frac{2k_B \gamma b \ln\frac{\bar{\omega}}{\bar{\omega}'}}{c} N^{\gamma-1}$$
$$+ \frac{b^2(1-2\gamma)}{cT} N^{2(\gamma-1)} + \frac{1}{2}\ln\frac{\bar{\omega}}{\bar{\omega}'} - \frac{k_B^2 T}{c}(\ln\frac{\bar{\omega}}{\bar{\omega}'})^2$$
（37）

$$\frac{\partial \ln W(T,N)}{\partial(\frac{1}{T})} = -T^2 F_T = \frac{bN^\gamma}{2k_B} - \frac{b^2}{c}N^{2\gamma-1} - \frac{T}{2} + \frac{k_B^2}{c}(\ln\frac{\bar{\omega}}{\bar{\omega}'})^2 NT^2$$
（38）

Eq (38) is in accordance with Luo-Lu's formula for the non-Arhenius temperature dependence of protein folding rate $k_f$[9] only with the exception of the symbol of $T/2$ term,

$$\frac{d\ln k_f}{d(\frac{1}{T})} = \frac{d\ln I_V}{d(\frac{1}{T})} = S + \frac{1}{2}T + RT^2 \tag{39}$$

where

$$S = \frac{\Delta E}{2k_B}(1-\frac{\Delta E}{\varepsilon}), \quad R = \frac{k_B}{2\varepsilon}\lambda^2 \tag{40}$$

$$\varepsilon = \bar{\omega}^2(\delta\theta)^2 \sum I_j \cong NI_0\bar{\omega}^2(\delta\theta)^2 \tag{41}$$

$$\lambda = \sum_j \ln \frac{\omega_j}{\omega'_j} \cong N \ln \frac{\bar{\omega}}{\bar{\omega}'} \qquad (42)$$

The temperature-sensitivity of the HM transition rate $W$ makes the histone modification level also temperature-dependent. Here we have predicted the nearly same non-Arhenius temperature dependence of the histone modification as usual protein folding. The experimental test of this theoretical prediction is awaited for.

3.3 The relation between histone modification rate $W$ and coherence length $N$ is deduced. The number of independent torsion angles in cooperative quantum transition is an important quantity in the theory. We call it coherence length. Eq (37) gives its relation to transition rate. The relation is dependent of parameter $\gamma$. For two limiting cases $\gamma=1$ ($\{\delta E_j\}$ summed up in phase) and 0 ($\{\delta E_j\}$ summed up out of phase) we have

$$\frac{\partial \ln W}{\partial N} = \frac{1}{2N} + \frac{b}{2k_B T} + \frac{1}{2} \ln \frac{\bar{\omega}}{\bar{\omega}'} - \frac{T}{c}(\frac{b}{T} + k_B \ln \frac{\bar{\omega}}{\bar{\omega}'})^2$$
$$\cong \frac{1}{2N} + \frac{b}{2k_B T} - \frac{b^2}{cT} = \frac{1}{N}(\frac{1}{2} + \frac{S}{T}) \qquad (\gamma=1) \qquad (43)$$
$$< 0$$

and

$$\frac{\partial \ln W}{\partial N} = \frac{1}{2N} + \frac{b^2}{cTN^2} + \frac{1}{2} \ln \frac{\bar{\omega}}{\bar{\omega}'} - \frac{k_B^2 T}{c}(\ln \frac{\bar{\omega}}{\bar{\omega}'})^2 \qquad (\gamma=0) \qquad (44)$$
$$\cong \frac{1}{2N}(1 + \frac{(\Delta E)^2}{k_B T \varepsilon}) > 0$$

respectively. The symbol of $\frac{\partial \ln W}{\partial N}$ is dependent of the parameter $\gamma$. As indicated in the previous work [9], for most proteins $S<0$ and $\left|\frac{S}{T}\right|$ is in the order of 100. So, for the case of $\gamma$ near 1 the transition rate decreases rapidly with coherence length. Through point mutation of some histone residue one can change the parameter $N$ and test the above relation between transition rate and coherence length.

3.4 In principle, quantum coherence can occur in all degrees of freedom of torsion angles nearby the binding sites of atomic groups. In addition to the torsion angles on histone chain the coherence may extend to some torsion angles on histone modifying enzyme. For example in yeast, the influence of H3K14ac on H3K4me3 was observed in experiment through SAGA complex of acetylation enzyme Gcn5 [14]. This can be explained by the quantum correlation existing in the torsion angles of the H3K14ac acetylation enzyme and protein domain Chdl chromodomain which links to histone H3 methylation.

3.5 The quantum nature of histone modification can be tested directly from the comparison between single HM transition and multi-site transition. In the above-mentioned bivalent model all torsion angles of histone chain between a pair of modifications are supposed to participate in the transition. However, the theory can be generalized to the case of single histone modification site. Suppose the polypeptide chain of length *N* neighboring to the single histone modification site participates in the transition cooperatively. Eq（26）and (30）describe the single HM transition as well but in this case $a^{(j)}_{\alpha'\alpha}$ should contain only one of two terms in Eq (10). As compared with single HM transition, the amplitude of pair transition is the summation of single transitions. So, a pair of histone modifications occurring not far from each other in the polypeptide chain can effectively strengthen the cooperative transition and enhance the transition rate. The superposition of probabilistic amplitude is characteristic of quantum mechanics. The experimental comparison between multi-site transition and single HM transition provides an additional clue to the test of the proposed quantum histone modification theory.

methylation with SAGA- and SLIK-dependent acetylation. Nature 2005, 433:434-438.